# A Survey on Peer-to-Peer and DHT

**Siamak Sarmady**
Grid Lab, School of Computer Science
Universiti Sains Malaysia, Penang, 11800, Malaysia.
sarmady@cs.usm.my

**Abstract –** *Distributed systems with different levels of dependence to central services have been designed and used during recent years. Pure peer-to-peer systems among distributed systems have no dependence on a central resource. DHT is one of the main techniques behind these systems resulting into failure tolerant systems which are also able to isolate continuous changes to the network to a small section of it and therefore making it possible to scale up such networks to millions of nodes. This survey takes a look at P2P in general and DHT algorithms and implementations in more detail.*

**Keywords:** Peer-to-Peer, DHT, Overlay Network, Hash table.

## 1   Introduction

Peer-to-Peer systems and applications have become popular in recent years. Perhaps the most commonly used forms of such applications are the file sharing and content distribution applications. Other applications of peer-to-peer systems have also been introduced which include higher quality voice and video applications. These new applications are gaining more attention because of the higher available bandwidth on the internet.

Failure tolerance of peer-to-peer systems is one of the most important aspects of this technology which can make it dependable. P2P systems are normally combined of several nodes. Because of the specific architecture and design, problem in some of the nodes can be isolated to a small section of the system, avoiding the disruption of the entire system. Huge amount of resources being shared in this way along with the failure tolerance of these systems promises a brilliant future for this technology.

One of the technologies being used in peer-to-peer systems to decrease dependability on central servers is distributed hash table or DHT. Napster's shut down case, not considering the reason, has something for us to learn. Systems which depend on a central service, (even if the main service is based on peer-to-peer technology) cannot be dependable because a problem with that central service might shut down the entire system. A similar problem due to the dependability of nodes to central authentication service caused shut down of the Skype for 2 days and put down most of the 220 million users [10][11].

DHT is an attempt to make peer-to-peer systems as independent as possible from any central service. In Napster's case a central server was used for resource discovery while DHT is trying to provide this in a completely distributed way.

In section 2 of this survey we will review the peer-to-peer technology. Section 3 will cover DHT concept and ideas behind it. In next section we will see some of the DHT algorithms. Section 5 will introduce some of the available toolkits and libraries being used to develop DHT applications. In the last section we will conclude our discussion.

## 2   Peer-to-Peer networks

Network systems are normally categorized as server or client according their rule in a specific service. In peer-to-peer systems nodes cannot easily be categorized as server or client because they might have server and client relationship with other nodes at the same time. For example file sharing application like Bittorrent might get parts of a file from a neighboring node while giving it parts of another file at the same time.

### 2.1   Main ideas and concepts

The main idea behind peer-to-peer systems has been to avoid a central resource which might become a bottleneck to the entire system. In the attempt of decentralization of systems, different applications have tried to distribute the system as much as possible and provide redundant servers for each service. In peer-to-peer architecture almost all of the nodes provide the same service. In fact similarity of nodes is the main characteristic of these systems.

Another main idea behind P2P is the concept of direct sharing of resource like bandwidth, CPU cycle, storage space and content among nodes without meditation of another server or central service [1]. In peer-to-peer systems nodes normally share similar resources. For example, in file sharing applications like Emule, the content or files are being shared. Skype shares the available bandwidth to make more high quality voice communications possible. Seti@Home is designed to shares the CPU processing cycle of nodes [9].

Due to independence from a central resource, P2P networks can scale up to thousands or even millions of nodes and the network has the self organizing capability in a way that addition or removal of a node will only cause a minimal change to the network.

## 2.2 Peer-to-Peer vs. Grid Computing

Despite having lots of similarities these two types of networks have important differences. In P2P networks the responsibility of management and maintenance of nodes is distributed to node users [1]. In comparison nodes of grid computing resources normally belong to organizations and the organization determines usage and sharing policies.

In addition computing resources in grid computing are more homogenous. In peer-to-peer systems nodes can operate on different types of hardware and even operating systems.

## 2.3 Pure P2P

Different applications categorized under the name peer-to-peer, have implemented ideas behind P2P in different levels. Napster used a central server as both the directory of node addresses and index of resources being hosted on those nodes. But the core service, namely the content distribution was provided by nodes. Shutdown of the central node due to legal cases resulted into complete shutdown of entire system. This case shows that a central service might cause the entire system to become useless. On the other hand if we can construct a peer-to-peer network without any kind of independence to a single node or a central service and at the same time maintain a good security in the system we will have a very reliable and dependable system.

In a pure P2P system, resources and functions are totally distributed and all nodes are completely equivalent in terms of their functionalities and tasks. Systems which consider some of the nodes as super nodes, or use a central authentication, index etc. services cannot be called pure P2P [1].

## 3 Distributed Hash Table

As we mentioned earlier resource sharing is one of the main ideas behind P2P networks. As the size of P2P networks increase, amount of shared resources like content increases. Finding a specific resource in P2P networks in a scalable manner is a real design challenge.

Traditional method used to find resources on the internet has been the name services like DNS. Hierarchical design of DNS distributes the load on several DNS servers and also guarantees that an available record will be found. The main problem of this approach is that the design is highly sensitive to failure or shutdown of root servers and servers near to root. In addition root servers might encounter scalability problems due to the high number of requests and central nature of the service. Early P2P content distribution networks were similarly relying on central index of the available resources on the network. Due to problems happened to such services, completely distributed approaches were adopted.

In this approach, all nodes are as important as others. Entries are distributed over the available nodes. A query searching for a resource is first being forwarded to local database. If it is not found in the local database it will be forwarded to other nodes. If the other node has the result of the query in its database, it will send back the result to the origin of the query; otherwise it will forward the query to its neighboring nodes. Sending the request to all the hosts or to many hosts on the network at the same time will flood and might collapse it. Gnutella used to handle queries in this way and because of the network load resulted from the Gnutella's flooding, the service become obsolete in favor of more efficient designs.

The new approach used the concept of overlay networks to limit the amount of queries being spread over the network. Each host in this method chooses a few of the nodes on the network as its close neighbors. Transferring queries is then limited to a few of the neighboring nodes. Routing of the queries is being performed in a more efficient way to be able to provide both the scalability and guarantee of finding the resource.

To be able to have a scalable solution, the approach should be able to isolate network changes to a small part of it, otherwise each small change will need update of the entire network and therefore a lot of overhead and bandwidth will be needed to converge it into a stable status again. P2P networks as mentioned earlier might contain thousands or millions of nodes so having a solution which can limit the changes to a very small part of the network is essential. Finding a resource from among those huge numbers of nodes needs a fast algorithm which can quickly point us to the place which hosts a resource.

Distributed Hash Table (DHT) is able to fulfill both of our requirements. DHT is a distributed data structure which holds key and value pairs in a completely distributed manner. DHT puts each key-value pair on a single node only (no replication). Each addition will only change data on a few nodes and the distribution of data is almost fair.

## 3.1 DHT basic concepts

As described earlier key-value pairs in DHT are distributed over the nodes of peer-to-peer network. To be able to determine on which node a specific pair should be stored we need a mapping method. In addition, adding and removing a node should not cause all the keys to be remapped. The mapping method being used in DHT is being called consistent hashing.

Consistent hashing divides the key space into partitions. This method normally uses the concept of distance to map a key to a specific node. Distance is a logical concept and should not necessarily be related to physical or network distance of nodes. In a P2P network a node which is physically in Germany could be nearer to a node in Canada than a node in the same country.

The mapping function is being used when we need to insert a new key-value pair into the hash table and also when we want to find the key. This function uses the key itself to determine the node which will store a pair. Then, when the same key is being queried, the same mapping function can determine the place which the key is being stored and therefore make retrieval of the value faster.

Using DHT each node when receiving a query tests to see if it has the key or not. If the key is not stored on the same node, the P2P application will use the mapping function to determine which of its neighbors has the least distance to the key-value's storing location. Then it forwards the query just to a few of its neighbors which are nearer to the storage location. Neighbors continue the same process until a node finds the key and sends it back.

Insert process happens in a same way. When a node receives an insert request, it uses the map function to see how near it is to the keys place. Then it tests to see if any of its neighbors is nearer to the place which key should be stored. If one of the neighbors is nearer to the place, then it passes the pair to that neighbor. The neighbor repeats the process and passes the pair to one of its neighbors, which is nearer to the place. This process continues until a node is not able to find a neighbor which is nearer to storing place than itself (according mapping function). At this moment that specific node understands that, it is the best place to store the key and stores the key in its local database.

In case one of the hosts goes down, the only effected key-value pairs are those which were stored on that specific node. From this moment another node will be nearer to position of the node which went down and therefore takes the responsibility of storing and responding to the queries near to that logical place on the network.

To avoid loss of stored keys some implementations try to replicate their database to only a few of the nearest neighbors from time to time. In this way, we yet have little storage need on each node (if we wanted to replicate the entire content of the nodes on every node it would become impractical) and also the key-value pairs stored on removed nodes are preserved [3][1][5].

## 4 DHT Algorithms

As described earlier, DHT is in abstract idea which helps us to reach our two main goals, complete independence from a central lookup server and tolerance to changes in the network. Different algorithms have been designed to implement DHT idea. Among the most popular implementations we can name Kademila [6], CAN [10], Tapestry [26], Pastry [27] and Chord [28]. In addition to those mentioned above there are other algorithms available which are less popular. We can mention Koorde [30], Viceroy [29], DKS [16], Accordion [22], Open-Hop [23] and Dipsea [15] among these. In this section we will have a brief review on techniques and design of some of the above mentioned algorithms.

### 4.1 Kademila

Kademila the same as most of other DHT algorithms is based on distance of nodes and key-value pairs. Each node in Kademila has a node ID. "SHA-1" hash algorithms is being used to build a 160-bit key from the node ID for each node. Distance between each two nodes is computed very easily by performing a XOR on keys of the two nodes. The resulting number of the XOR operator will be used as the distance between two nodes. Each node has more knowledge about nearer nodes and less knowledge about far away nodes, meaning that database of nodes contains more entries of near neighbors. Storage place of each key-value pair is also determined by distance. Each pair is stored on the nearest node to it. Distance of a key-value pair to a node is calculated the same way we calculate the distance of two nodes from each other. A hash from the key (of key-value pair) is used to calculate distance of a pair to a node. An insert query is first being sent to a few of neighbors of the initiating node and then travels to the node which is the nearest node to the hash of the key-value pair. If a node goes down, another node will take the responsibility of that region of the hash table space.

Kademila uses parallel, asynchronous queries to avoid timeout delays from failed nodes. Kademila uses 128bit routing table to speed up the search and maintains a separate list for each bit. Every list belongs to a specific distance from current node. As a result of this efficient method, each search iteration in a network with $2^n$ nodes will take at most n steps to complete [6][12].

## 4.2 Chord

Chord as another DHT algorithm has similarities to Kademila in that, it also uses "SHA-1" hash function as a base for consistent hashing. Nodes in Chord are placed on a Ring. Both node IDs and keys (hash from key-value pairs) are placed on the same ring. The hash function produces an m-bit identifier for both nodes and keys for this purpose. Each node has a successor and predecessor. Insertion of a new node between two older nodes involves the update of successor of one of those node and predecessor of the other. Adding a key-value involves a same method as Kademila. A key is assigned to the first node whose identifier is equal or follows the identifier of the key.

If a node goes down, after a short period successor and predecessor nodes will discover this using their Ping function and will update their successor and predecessor pointers. Responsibility of the key-space of the failed node will be transferred to the other nodes. Because each node in chord transfers queries to only one successor node, the flooding problem of the older resource discovery methods does not happen. [28]

## 5  DHT Libraries

Different groups and individual have created DHT libraries based on earlier mentioned algorithms or with small differences to those basic algorithms. In this section we will list a few of the available libraries.

Overlay Weaver [14] toolkit provides multiple routing algorithms, Chord, Kademlia, Koorde, Pastry and Tapestry. It is also possible to implement new algorithms using this toolkit. JDHT [17] implements the DKS algorithm. GISP provides another distributed hash table library and can be used with JXTA toolkit [18]. Bamboo DHT implements the Pastry algorithm [19]. OpenChord implements Chord algorithm [20]. Jxta-meteor is a platform to develop distributed hash tables and Chord algorithm has been implemented on it as an example [21]. Libdht [24] provides a DHT library for Donkey Kademila implementation. NUPastery library provides an implementation of Pastry algorithm [25].

Author has been able to find other libraries which are less popular and therefore only the earlier mentioned libraries and toolkits are listed.

### 5.1 API

Though algorithm and operation of different DHT methods differs very much from each other, because they implement a single abstract idea of a Hash Table, they have almost the same API as normal undistributed hash tables. Generally these algorithms provide an initialize method to maintain initialization of the communication with other nodes and preparation of local database. Then two most important functions are PutValue (key) and GetValue (key). The first function is responsible for adding a new key-pair value to the distributed collection. The other function queries the distributed data structure for a specific key-value pair. In background, these functions might forward the requests to other hosts but what a programmer sees from the library interface is normally transparent. A Remove (key) function is sometimes available which will find and remove the instances of a specific key in the distributed database.

Different algorithms have their specific internal messages and functions which differ based on the design. For example a Ping and Pong (response to ping) is implemented to allow verifying that a node is alive. As another example, a function can be provided to replicate the keys to near nodes before stopping a specific node.

## 6  Conclusions

We had a brief review on Peer-to-Peer distributed systems and discussed different types of software which is being referred under this name. Advantages and weaknesses of Peer-to-Peer systems were discussed and role of DHT in complete distribution of resource discovery and failure tolerance was reviewed. At the end we presented some of the designed DHT algorithms followed by a list of available libraries and toolkits.